# Continuous-wave operation at 30 MV/m of a superconducting radio-frequency gun cryomodule with an exchangeable cathode plug

T. Konomi,[1,*] T. Xu,[1,†] Y. Choi,[1] K. Elliott,[1] B. Gower,[1] B. Tousignant,[1] J. Wenstrom,[1] A. Taylor,[1] C. Compton,[1] W. Chang,[1] X. Du,[1] W. Hartung,[1] S.-H. Kim,[1] S. Miller,[1] D. Morris,[1] M. Patil,[1] L. Popielarski,[1] J. Wei,[1] Z. Yin,[1] J. Smedley,[2] J. Maniscalco,[2] C. Adolphsen,[2] M. Murphy,[2] D. A. Gonnella,[2] M. P. Kelly,[3] T. Petersen,[3] A. Arnold,[4] S. Gatzmaga,[4] R. Steinbrück,[4] P. Murcek,[4] R. Xiang,[4] J. Teichert,[4] and J. Lewellen[5]

[1]Facility for Rare Isotope Beams, Michigan State University, East Lansing, MI 48824, USA
[2]SLAC National Accelerator Laboratory, Menlo Park, CA 94025, USA
[3]Argonne National Laboratory, Lemont, IL 60439, USA
[4]Helmholtz-Zentrum Dresden-Rossendorf e.V., D-01328 Dresden, Germany
[5]Los Alamos National Laboratory, Los Alamos, NM 87545, USA

## ABSTRACT

A prototype 185.7 MHz superconducting radio-frequency electron gun cryomodule has been developed for the proposed LCLS-II-HE Low Emittance Injector at SLAC. It has an in-situ ultra-high-vacuum exchangeable cathode plug system compatible with semiconductor photocathodes. With a copper plug, the gun achieved continuous-wave RF operation at a cathode field of 30 MV/m for 1.5 hours. During full-field operation, the pressure remained below $1\times10^{-10}$ Torr; and the intrinsic quality factor was $2.4\times10^{9}$ at the design stored energy. Complete plug removal and reinsertion were demonstrated with the cavity at 4.4 K. During the exchange, the pressure remained below $1\times10^{-9}$ Torr except for brief transients associated with gate-valve operation. No measurable degradation in high-field cavity performance and no systematic increase in field-emission-induced X-rays were observed after plug reinsertion; the pre-exchange high-field conditions were re-established without additional multipacting conditioning. These results represent a significant improvement over past continuous-wave SRF gun systems with exchangeable plugs.

## I. INTRODUCTION

The Linac Coherent Light Source (LCLS) at SLAC is being upgraded through the LCLS-II High Energy project (LCLS-II-HE), which will increase the energy of the continuous-wave (CW) LCLS superconducting linac (LCLS-SC) from 4 to 8 GeV, extending its hard X-ray photon-energy reach to approximately 13 keV. To further extend this reach to 20 keV, SLAC initiated the development of a low-emittance injector (LEI) that uses a superconducting radio-frequency (SRF) gun in a standalone cryomodule [1]. The LEI design targets a reduction in electron beam emittance by more than a factor of two. A cathode field of 30 MV/m accelerates a bunch rapidly to suppress space-charge-induced emittance growth. With optimized emittance compensation and velocity compression, the LEI would produce 100 pC bunches of length 1 mm rms with 100 nm emittance at 100 MeV, where the emittance is fairly insensitive to the cathode gradient in the 25–35 MV/m range [2]. This field is substantially higher than the cathode field of about 18 MV/m used with the existing normal-conducting radio-frequency (NC-RF) gun for LCLS-SC [3].

More broadly, high-brightness CW electron sources are important for next-generation X-ray free-electron lasers (XFELs), energy-recovery linacs, and other high-duty-factor accelerators [4]. NC-RF guns have produced high-brightness electron beams and can support CW or high-duty-factor operation. However, the RF power dissipation makes it difficult to substantially increase the CW cathode field. An SRF gun offers the complementary advantages of high CW cathode field, low RF dissipation, and an ultra-high-vacuum (UHV) environment favorable for long photocathode lifetimes [5]. These characteristics make SRF guns promising electron sources for high-brightness, high-repetition-rate accelerators.

*Contact author: konomi@frib.msu.edu
† Contact author: XuTi@frib.msu.edu

Operating a high-performance photocathode in an SRF gun presents several challenges. Photocathodes with high quantum efficiency (QE) and low mean transverse energy (MTE) generally require preparation in a dedicated system, transport under UHV, and periodic replacement [6]. An exchangeable normal-conducting cathode plug must be inserted into the high-electric-field region of an SRF cavity without introducing particulate contamination that could increase field emission (FE) or degrade the SRF cavity performance. The plug support assembly, referred to as the cathode stalk, must limit RF leakage from the cavity, limit heat leakage into the cavity, provide reproducible alignment and reliable electrical contact, and allow the plug temperature to be controlled independently of the cavity temperature. As multipacting (MP, resonant build-up of electrons via secondary emission and acceleration by the RF field) in the cavity, fundamental power coupler, or stalk may release residual gas and poison a high-QE photocathode [7], mitigation of MP is very important. This can be done by careful design of the cavity and application of DC bias to the coupler antenna and stalk, though this further complicates the system design.

Existing SRF guns have demonstrated important subsets of the desired capabilities, including high-current CW beam production, operation with high-QE photocathodes, plug exchange, and high-gradient cavity performance [8-14]. However, simultaneously achieving a high CW cathode field, UHV plug exchange without cavity warming, and preservation of high-field SRF performance remains challenging. In particular, cathode plug insertion carries the risk of introducing particulates that increase FE, while, as described above, the stalk design requirements are challenging.

A 185.7 MHz quarter-wave-resonator (QWR) SRF gun was developed as a prototype for the LCLS-II-HE LEI with the goal of addressing the challenges outlined above [15-17]. The SRF gun system was qualified through staged tests of major subsystems before full system assembly and testing [18]. We report on CW operation of the full SRF gun cryomodule at a cathode field of 30 MV/m using an exchangeable copper plug, along with demonstration of complete plug removal and reinsertion under UHV with the cavity maintained at 4.4 K and without measurable degradation of RF performance or increase in FE-induced X-rays.

## II. SRF GUN DESIGN AND VALIDATION

As shown in Fig. 1, the gun cavity is a QWR with a 70 mm accelerating gap. The cavity operates at the 7th subharmonic of the LCLS-II linac frequency (1.3 GHz) and is designed to produce a cathode field ($E_c$) of 30 MV/m [19]. The low operating frequency allows for long bunches with low peak current, which can then be shortened by velocity compression [2]. The low frequency yields a lower surface resistance, allowing for operation at 4.4 K instead of the sub-atmospheric 2 K operation typically used for higher frequencies such as 1.3 GHz. The cavity design was oriented toward MP suppression and minimization of the peak surface electric field to reduce dark current [20-22]. Elliptically-rounded corners for the cavity's cathode port and the plug help to reduce the local electric-field enhancement while the cavity's rounded "anode dome" geometry weakens low-field MP barriers. For the as-designed cavity shape, the maximum surface electric field ($E_{peak}$) is 34 MV/m and is located on the niobium cavity surface away from the cathode port as seen in Fig. 1b. For the as-fabricated cavity after surface preparation, the cathode port inner diameter was larger than designed; at the design field with the plug in its nominal flush position, a model based on measurements of the prototype cavity predicts $E_{peak}$ = 36 MV/m at the cathode plug edge, giving $E_{peak}/E_c$ = 1.20 (with little change in the peak magnetic field $B_{peak}$). Thus, though the local field enhancement for the prototype cavity is worse than desired, $E_{peak}/E_c$ remains relatively modest. Tighter control of the cathode-port dimensions during fabrication and surface preparation may help to reduce the local field enhancement for future cavities.

To accommodate the complex QWR geometry, surface processing requirements were incorporated into the cavity design. The cavity has four ports (two of which can be seen in Fig. 1) for electropolishing (EP) electrodes and high-pressure rinsing (HPR) with ultra-pure water, which are critical surface preparation steps [23]. Including the beam port, five access ports are available for HPR. A ray-tracing simulation predicts that the water jets can reach 99.4% of the internal cavity surfaces. A full Nb "blank" cavity without a cathode port was built to validate the cavity performance without the complication of the insertable plug. The blank cavity reached $E_c$ = 30 MV/m in a cold test without measurable FE-induced X-rays [24].

The plug is inserted via a transfer rod into a spring-locked socket in the cathode stalk assembly and then the rod is retracted. The plug and stalk are thermally and electrically isolated from the SRF cavity [25, 26]. The stalk position is adjustable to precisely align the cathode tip to be flush with the cavity wall and centered in the cathode port. The stalk also provides internal reference surfaces for the automated plug insertion system (Section III). The stalk is made of stainless steel to limit conduction heat transfer to the 4.4 K cold mass; it is plated with 20 μm of copper to reduce RF dissipation and with 1 μm of gold to reduce thermal radiation. An independent cooling circuit allows the plug temperature to be controlled from approximately 35 K to 300 K while the cavity remains at 4.4 K, a capability which has not been previously demonstrated in an SRF gun system. Ceramic interfaces electrically isolate the stalk; they are coated with TiN to reduce secondary electron emission. A high voltage (HV) bias of up to ±5 kV can be applied to suppress MP in the coaxial region of the stalk [27]. RF power coupled into the stalk region is attenuated before reaching the HV supply by a strip-line stub filter (Section IV). In standalone RF tests, the stalk assembly reached surface fields compatible with $E_c$ = 30 MV/m, and MP in the stalk was suppressed with a bias of approximately 400 V.

The SRF gun cryomodule incorporates a novel, compact, superconducting solenoid doublet with independently-powered windings, allowing the focusing strength and magnetic centroid to be adjusted for optimized emittance compensation [28]. The iron-free design eliminates magnetic hysteresis, providing reproducible magnetic fields without standardization procedures and a true zero field when the solenoid is off. Additional coils are included to correct field asymmetries in the integrated gun system and provide beam steering. The cavity, stalk, fundamental power coupler, tuner, and magnet package are integrated into a "bottom-up" cryomodule configuration [16], whose features are shown schematically in Fig. 1a.

Subsystems were qualified through staged tests before final assembly of the full cryomodule. These tests included RF and thermal tests of the stalk; particle-generation measurements and alignment tests of the cathode insertion system [29-31]; cryomodule operation with the blank cavity; and cold testing of the full cavity (described below). This approach allowed the RF, vacuum, thermal, alignment, and cryogenic performance of the principal subsystems to be validated and provided opportunities to make corrections before full system testing. The principal parameters of the SRF gun cavity are listed in Table I.

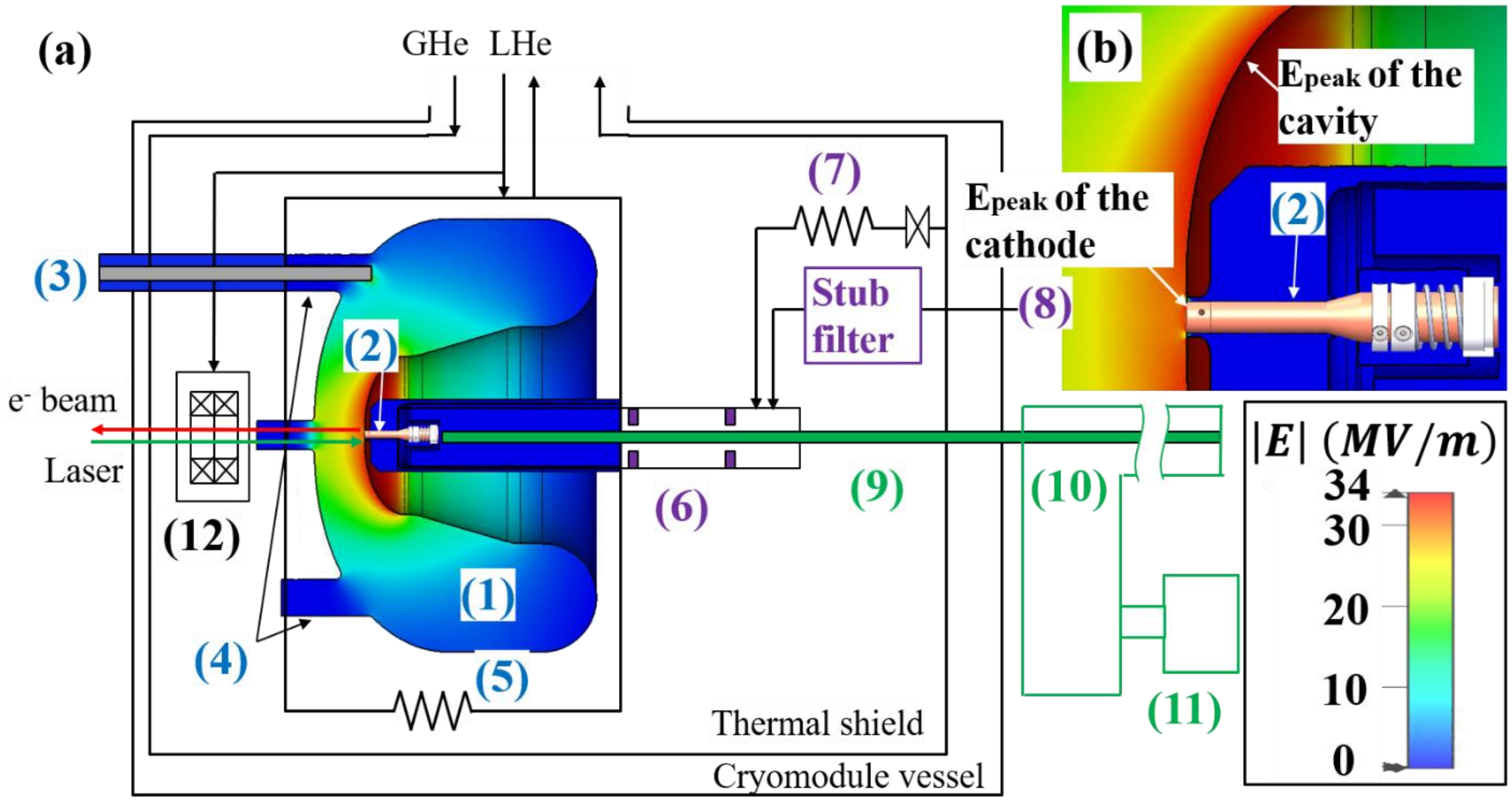

FIG. 1. (a) Schematic of the SRF gun cryomodule and cathode insertion system: (1) SRF cavity, (2) cathode plug, (3) fundamental power coupler, (4) rinse ports for EP and HPR, (5) frequency tuner, (6) cathode stalk with alignment-sensor rings, (7) cathode heater, (8) cathode bias line with stub filter, (9) transfer rod, (10) automated alignment system, (11) UHV suitcase, (12) superconducting solenoid package. The cathode plug is transported via the transfer rod to its operating position in the cathode stalk. (b) Enlarged view of the cathode region showing the plug and the location of $E_{peak}$ at the plug edge and the maximum electric field on the niobium cavity surface. The color map represents the RF electric field magnitude.

Prior to cryomodule assembly, the full SRF gun cavity was Dewar-tested without the cathode plug. Figure 2 shows the intrinsic quality factor ($Q_0$) and field-emission-induced X-rays as a function of the RF field measured during the fourth cold test. The cavity reached a field equivalent to $E_c$ =30 MV/m (meaning that the RF stored energy in the test without the cathode plug was equal to the stored energy needed to reach $E_c$ = 30 MV/m with the cathode in the flush position), while the FE-induced X-rays remained at the background level throughout the measurement.

The cathode-port geometry was designed to minimize transverse electric fields at the cathode surface with the plug in the flush-with-the-wall position. In this plug position, the plug shields the sharp port edge from the RF field as well, suppressing the local field enhancement produced without the plug. Since the plug was absent during the cold test, the peak surface electric field was located at the sharp cathode-port edge: $E_{peak}$ was approximately 46.5 MV/m at $E_c$ = 30 MV/m, compared with $E_{peak}$ = 36 MV/m with the plug in the flush position, as described above.

TABLE I. Principal design parameters and operating goals for the SRF gun cavity. $Q_0$ = intrinsic quality factor; $R_a$ = shunt impedance (linac definition, calculated with beam velocity = light speed).

| Design Parameter | |
|---|---|
| RF frequency | 185.7 MHz |
| $R_a/Q_0$ | 113 Ω |
| Geometry factor | 84.5 Ω |
| **Operation Goals** | |
| Cathode field ($E_c$) | 30 MV/m |
| $E_{peak}$ (design) | 34 MV/m |
| $E_{peak}$ (as-fabricated) | 36 MV/m |
| $B_{peak}$ | 53 mT |
| Stored energy | 21.4 J |
| Final electron energy | 1.8 MeV |
| Operating temperature | 4.4 K |
| $Q_0$ at $E_c$ = 30 MV/m | $2.4\times10^9$ |

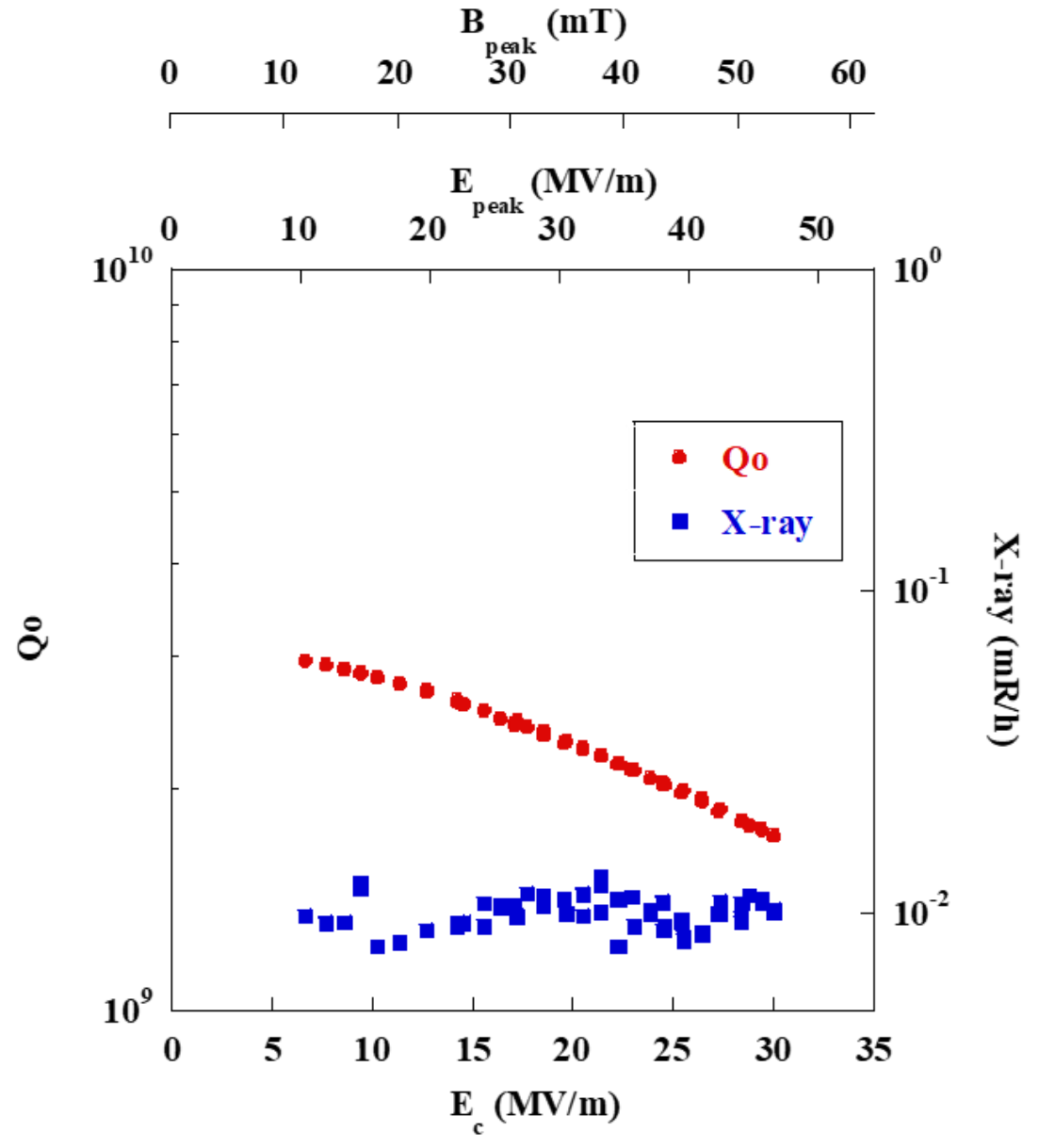

FIG. 2. Intrinsic quality factor $Q_0$ and X-ray dose rate measured with the SRF gun cavity at 4.4 K, without the cathode plug. The horizontal axes show the peak surface electric field ($E_{peak}$), the peak surface magnetic field ($B_{peak}$), and the equivalent cathode field ($E_c$).

## III. CATHODE PLUG EXCHANGE AT 4.4 K

The cathode insertion system (CIS), developed by the team at Helmholtz-Zentrum Dresden-Rossendorf (HZDR) based on their experience with previous cathode transfer systems for SRF guns at HZDR and Helmholtz-Zentrum Berlin (HZB) [32, 33], transports the plug from a UHV transport "suitcase" to its operating position in the stalk as shown in Fig. 3. The CIS is based on the common insert design and similar transfer mechanisms developed at HZDR and HZB, although the two systems differ in their cathode exchange configurations. The shared design offers the potential for collaborative development of high performance photocathodes and cross-facility testing of different cathode preparation technologies. The longer travel distance of 1.37 m required for our low-frequency QWR-based system makes precise and particle-free insertion more challenging. To address this, a key new feature of the CIS is automated alignment of the long transfer rod during insertion. The insertion system relies on two sets of contact electrodes on the transfer rod that sense two internal rings aligned with the stalk axis. Contact with the rings allows computer-controlled stages to correct the transfer rod trajectory, which accounts for the rod sag. This active alignment provides a transverse positioning reproducibility better than 100 µm and allows for plug exchange without unintended mechanical contact [29-31], thereby minimizing the risk of particle contamination of the cavity.

Before plug exchange, the full SRF gun cryomodule was operated in CW for more than 3 hours at a stored energy of 21.4 J, with a copper plug recessed by approximately 2 mm from the flush position. In this position, the cathode tip surface field is 19.7 MV/m and the peak electric field on the cavity wall is approximately 42 MV/m (as seen in Table I, this is the design stored energy, corresponding to $E_c$ = 30 MV/m if the plug tip is flush with the cathode opening). Only a low level of FE X-rays (< 0.05 mR/hr at a distance of 1 m) was observed in this condition. A calorimetric measurement yielded $Q_0 = 2.4\times10^9$, consistent with the standalone cold test of the cavity. These measurements established the cavity's pre-exchange performance baseline.

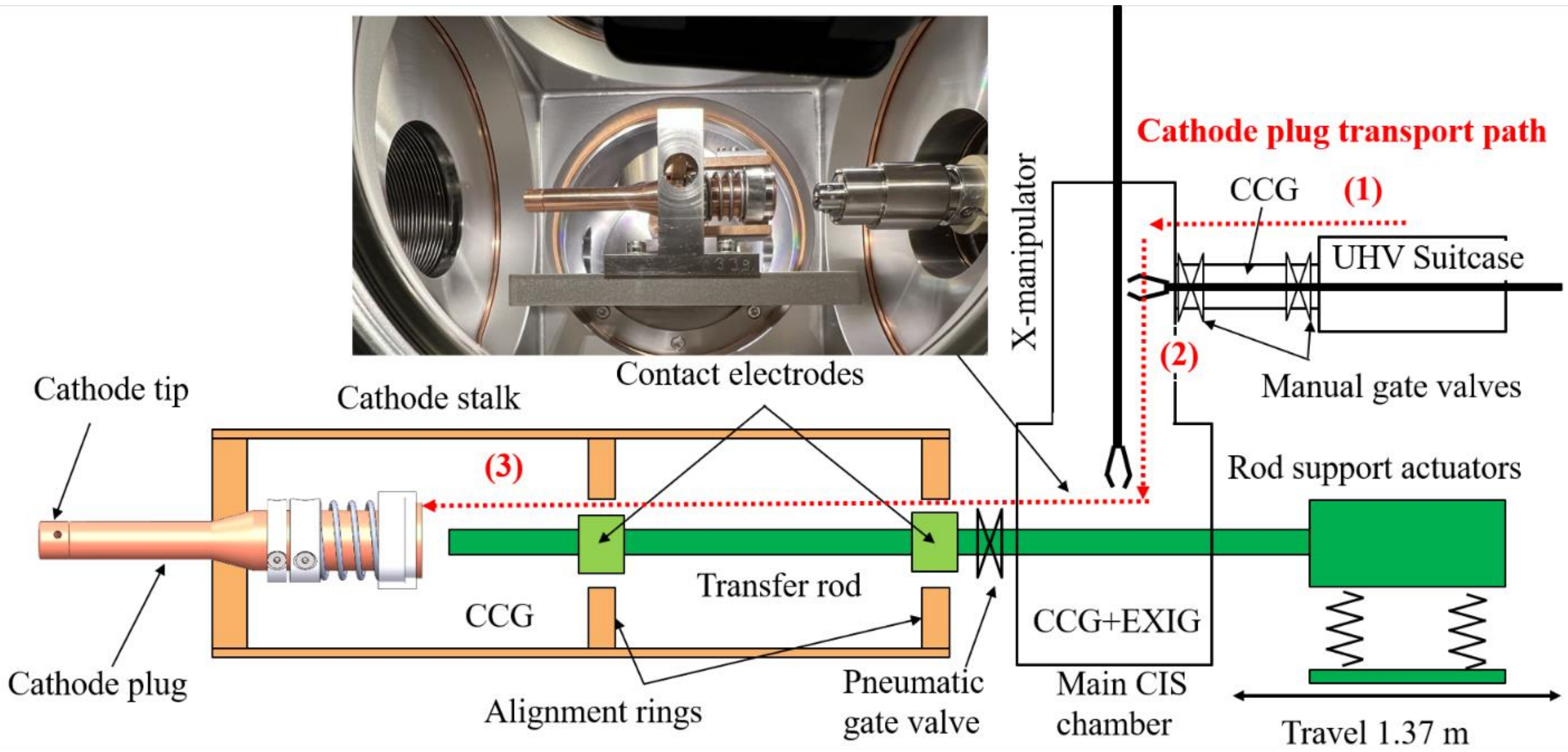

FIG. 3. Schematic of the cathode insertion system (CIS) and cathode plug transfer path. The red dashed arrows indicate the transfer sequence: (1) transport from the UHV suitcase, (2) transfer to the insertion axis, and (3) insertion into operating position. Two sets of contact electrodes help to determine the stalk axis position and angle, allowing for trajectory corrections over the 1.37 m insertion path. The system includes two manually-operated gate valves and one pneumatically-operated gate valve. The pressure is measured via a cold cathode gauge (CCG) connected to the cathode stalk and the cavity, a CCG and an extractor ion gauge (EXIG) on the main CIS chamber, and a CCG on the vacuum region between the main CIS chamber and UHV suitcase.

The copper plug was then moved from the cavity to the UHV suitcase (progressing from Location 3, through 2, to 1 in Fig. 3), and subsequently returned to its operating position (progressing from Location 1 to 3), while the cavity remained at 4.4 K. Between gate valve operations during cathode exchange, the measured pressures remained $< 1 \times 10^{-9}$ Torr. Brief pressure increases were observed during gate valve operation, reaching up to $2 \times 10^{-8}$ Torr in the main CIS chamber and $6 \times 10^{-9}$ Torr on the cavity side. After plug reinsertion, the cavity returned to the pre-exchange high-field condition; additional MP conditioning was not needed. No systematic increase in FE X-rays was observed, demonstrating that the plug exchange was done without detectable particle-induced degradation of the SRF cavity.

After cathode insertion, the room-temperature plug was cooled through contact with the cathode stalk, which was in turn cooled by helium gas at 30–40 K with a mass flow rate of 0.07 g/s. The cooling gas outlet temperature showed a transient increase upon plug insertion, followed by a decay with a characteristic time of approximately 45 s. Although the plug temperature was not directly measured, no noticeable transverse displacement was observed by a camera viewing the plug along the beam axis. Repeated cathode insertion and removal at 4.4 K showed a reproducible cavity resonant frequency, indicating consistent plug positioning at the level of measurement sensitivity.

## IV. RF-HEATING MITIGATION AND FULL-FIELD CW OPERATION

Following the successful plug exchange test, the performance of the full SRF gun cryomodule was evaluated with the plug in its nominal flush position. An earlier inductive filter for mitigation of RF leakage to the HV supply had produced localized RF heating and an associated pressure excursion. The inductive filter was replaced by a more robust strip-line stub filter. The stub length was tuned to minimize the RF magnetic field near the HV feedthrough at 185.7 MHz, thereby reducing local surface currents and RF dissipation. The copper strip-line was clamped between 99.5% alumina plates, providing electrical insulation and improved thermal anchoring.

The cavity with the copper plug in its nominal flush position was operated in CW for 1.5 hours at the design field of $E_c = 30$ MV/m with the stub filter. No significant RF-induced heating of the bias feedthrough was observed, and the pressure remained below $1\times10^{-10}$ Torr (the low limit of the pressure gauge). The FE X-ray level for $E_c = 30$ MV/m was approximately 1 mR/hr at a 1 m distance. This level was higher than that measured during the earlier plug exchange test and appeared after the pressure excursion associated with the failure of the inductive filter. Additional testing was done with the plug completely removed, producing a comparable X-ray level, which indicates that the plug was not the dominant source of FE X-rays.

These results, along with the plug exchange test discussed in Section III, demonstrate high-field CW operation of an SRF gun cryomodule with an exchangeable normal-conducting plug. Stable vacuum was maintained during full-field CW operation. During the test, the cooling-line temperatures for the thermally-isolated plug and stalk were approximately 45 K. In a separate thermal control test, the plug was heated to approximately 300 K while the cavity remained at 4.4 K.

Figure 4 compares the reported peak on-axis RF electric fields achieved during CW operation for SRF and NC-RF gun systems with in-situ exchangeable cathode plugs [8, 10, 11, 34-42]. The peak on-axis field was selected to provide a more consistent comparison among guns with different cavity geometries and cathode positions. The demonstrated high-field CW operation with a copper plug and successful plug exchange provides a promising basis for subsequent operation with high-QE, low-MTE photocathodes, as needed for the LCLS-II-HE Low Emittance Injector.

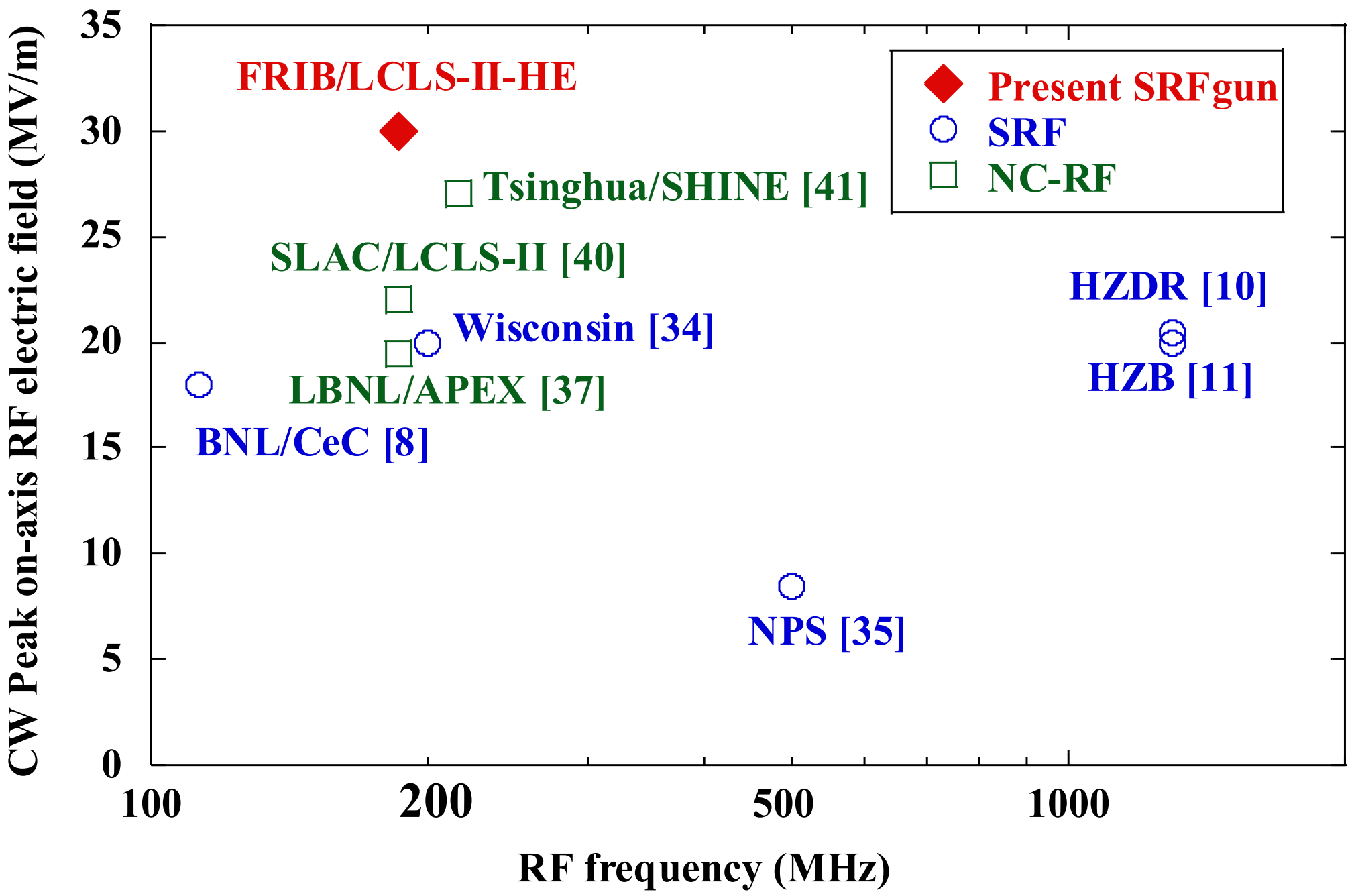

FIG. 4. Reported peak on-axis RF electric fields achieved during CW operation as a function of RF frequency for SRF (circles) and NC-RF (squares) electron guns with exchangeable cathodes. The red diamond represents the LCLS-II-HE SRF gun developed at FRIB. Values were taken from published results or estimated from published axial field distributions where necessary; the references are included for all values from the literature.

## V. CONCLUSION AND OUTLOOK

We have developed a full 185.7 MHz superconducting radio-frequency gun cryomodule and demonstrated operation at a high continuous-wave cathode field with a copper cathode plug. Complete plug removal and reinsertion were performed with the cavity at 4.4 K, with no measurable degradation of the high-field cavity performance and no increase in field emission attributable to the plug exchange. The cryomodule was operated in CW for 1.5 hours with the plug in its nominal flush position and a cathode field of 30 MV/m. These results demonstrate that plug exchange, UHV cleanliness, and high-field CW SRF operation can be simultaneously maintained in a full SRF gun cryomodule.

These results open the door to future operation of the SRF gun cryomodule with high-performance, low-mean-transverse-energy photocathodes for the LCLS-II-HE Low Emittance Injector. More broadly, this work addresses a central challenge for the SRF gun community: combining the high-duty-factor and high-gradient capability of an SRF cavity with the flexibility to prepare, transport, exchange, and operate high-performance photocathodes. Our development efforts advance the state-of-the-art and point toward a practical path to replaceable photocathodes for future high-brightness, high-average-current injectors.

## ACKNOWLEDGMENTS

This work was supported by the U.S. Department of Energy, Office of Science, under Contract No. DE-AC02-76SF00515.

[1] Linac Coherent Light Source, LCLS Strategic Facility Development Plan (SLAC National Accelerator Laboratory, 2020), Sec. 5.3, pp. 55–56. https://lcls.slac.stanford.edu/sites/default/files/2023-10/LCLS_Strategic_Development_Plan.pdf

[2] F. Ji, C. E. Adolphsen, C. Mayes, T. O. Raubenheimer, R. Coy, L. Xiao, L. Ge, and J. Qiang, "Beam Dynamics Studies on a Low Emittance Injector for LCLS-II-HE," in Proc. 5th North American Part. Accel. Conf. (NAPAC 2022), Albuquerque, NM, USA

(JACoW, 2022), pp. 619–622, Paper WEPA02. 10.18429/JACoW-NAPAC2022-WEPA02

[3] F. Zhou, C. Adolphsen, D. Dowell, and R. Xiang, "Overview of CW Electron Guns and LCLS-II RF Gun Performance," Front. Phys. **11**, 1150809 (2023). 10.3389/fphy.2023.1150809

[4] C. Adolphsen *et al.*, European Strategy for Particle Physics – Accelerator R&D Roadmap, CERN Yellow Rep. Monogr., CERN-2022-001 (2022). 10.23731/CYRM-2022-001

[5] R. Xiang, A. Arnold, and J. W. Lewellen, "Superconducting radio frequency photo-injectors for CW-XFEL," Front. Phys. **11**, 1166179 (2023). 10.3389/fphy.2023.1166179

[6] R. Xiang and J. Schaber, "Review of Recent Progress on Advanced Photocathodes for Superconducting RF Guns," Micromachines **13**, 1241 (2022). 10.3390/mi13081241

[7] I. Petrushina, V. N. Litvinenko, I. Pinayev, K. Smith, G. Narayan, and F. Severino, "Mitigation of multipacting in 113 MHz superconducting rf photoinjector," Phys. Rev. Accel. Beams **21**, 082001 (2018). 10.1103/PhysRevAccelBeams.21.082001

[8] I. Petrushina, V. N. Litvinenko, Y. Jing, J. Ma, I. Pinayev, K. Shih, G. Wang, Y. H. Wu, Z. Altinbas, J. C. Brutus, *et al.*, "High-brightness continuous-wave electron beams from a superconducting radio-frequency photo-emission gun," Phys. Rev. Lett. **124**, 244801 (2020). 10.1103/PhysRevLett.124.244801

[9] E. Wang, V. N. Litvinenko, I. Pinayev, M. Gaowei, J. Skaritka, S. Belomestnykh, I. Ben-Zvi, J. C. Brutus, Y. Jing, J. Biswas, *et al.*, "Long lifetime of bialkali photocathodes operating in high-gradient superconducting radio-frequency gun," Sci. Rep. **11**, 4477 (2021). 10.1038/s41598-021-83997-1

[10] J. Teichert, A. Arnold, G. Ciovati, J.-C. Deinert, P. Evtushenko, M. Justus, J. M. Klopf, P. Kneisel, S. Kovalev, M. Kuntzsch, *et al.*, "Successful user operation of a superconducting radio-frequency photoelectron gun with Mg cathodes," Phys. Rev. Accel. Beams **24**, 033401 (2021). 10.1103/PhysRevAccelBeams.24.033401

[11] T. Kamps *et al.*, "First beam commissioning of the HZB SRF photoelectron gun," in Proc. 16th Int. Part. Accel. Conf. (IPAC 2025), Taipei, Taiwan (JACoW, 2025), pp. 1718–1721, Paper WECN2. 10.18429/JACoW-IPAC2025-WECN2

[12] H. Jia, T. Li, T. Wang, Y. Zhao, X. Zhang, H. Xu, Z. Liu, J. Liu, L. Lin, H. Xie, *et al.*, "High-brightness megahertz-rate beam from a direct-current and superconducting radio-frequency combined photocathode gun," Phys. Rev. Res. **6**, 043165 (2024). 10.1103/PhysRevResearch.6.043165

[13] T. Konomi, K. Hara, Y. Honda, K. Hosoyama, H. Inoue, E. Kako, Y. Kondo, M. Masuzawa, M. Omet, T. Takatomi, *et al.*, "Horizontal test results of 1.3 GHz superconducting RF Gun #2 at KEK," in Proc. 21st Int. Conf. RF Supercond. (SRF 2023), Grand Rapids, MI, USA, (JACoW, 2023), pp. 540–544, Paper TUPTB049. 10.18429/JACoW-SRF2023-TUPTB049

[14] E. Vogel *et al.*, "High gradients at SRF photoinjector cavities with low RRR copper cathode plug screwed to the cavity back wall," arXiv:2310.02974 (2023). 10.48550/arXiv.2310.02974

[15] J. W. Lewellen, C. Adolphsen, R. Coy, L. Ge, F. Ji, M. J. Murphy, L. Xiao, A. Arnold, S. Gatzmaga, P. Murcek, *et al.*, "Status of the SLAC/MSU SRF gun development project," in Proc. 5th North American Part. Accel. Conf. (NAPAC 2022), Albuquerque, NM, USA (JACoW, 2022), pp. 623–626, Paper WEPA03. 10.18429/JACoW-NAPAC2022-WEPA03

[16] S. J. Miller, C. Adolphsen, A. Arnold, Y. Choi, C. Compton, R. Coy, X. Du, L. Ge, D. B. Greene, W. Hartung, *et al.*, "Status of the SLAC/MSU SRF gun development project," in Proc. 21st Int. Conf. RF Supercond. (SRF 2023), Grand Rapids, MI, USA (JACoW, 2023), Paper FRIBA07. 10.18429/JACoW-SRF2023-FRIBA07

[17] "LCLS-II-HE Low Emittance Injector Conceptual Design Report," SLAC Report LCLSIIHE-1.1-DR-0418-R0 (Mar. 2022).

[18] T. Xu *et al.*, "Low-emittance SRF photo-injector prototype cryomodule for the LCLS-II high-energy upgrade: design and fabrication," in Proc. Int. Part. Accel. Conf. (IPAC 2023), Venice, Italy, (JACoW, 2023), pp. 1396–1399, Paper TUPA028. 10.18429/jacow-ipac2023-tupa028

[19] S. H. Kim, W. Hartung, T. Konomi, S. J. Miller, M. S. Patil, J. T. Popielarski, K. Saito, T. Xu, C. Adolphsen, L. Ge, *et al.*, "Design of a 185.7 MHz superconducting RF photoinjector quarter-wave resonator for the LCLS-II-HE low emittance injector," in Proc. 5th North American Part. Accel. Conf. (NAPAC 2022), Albuquerque, NM, USA (JACoW, 2022), pp. 245–248, Paper MOPA85. 10.18429/JACoW-NAPAC2022-MOPA85

[20] Z. A. Conway, B. M. Guilfoyle, H. Guo, M. Kedzie, M. P. Kelly, and T. Reid, "Progress toward 2 K high-performance half-wave

resonators and cryomodule," in Proc. 18th Int. Conf. RF Supercond. (SRF 2017), Lanzhou, China (JACoW, 2018), pp. 692–694, Paper WEYA05. 10.18429/JACoW-SRF2017-WEYA05

[21] T. Xu *et al.*, "Completion of FRIB superconducting linac and phased beam commissioning," in Proc. 20th Int. Conf. RF Supercond. (SRF 2021), Virtual Conference (JACoW, 2022), p. 197, Paper MOOFAV10. 10.18429/JACoW-SRF2021-MOOFAV10

[22] S. H. Kim, W. Chang, W. Hartung, J. T. Popielarski, and T. Xu, "Conditioning of low-field multipacting barriers in superconducting quarter-wave resonators," in Proc. 5th North American Part. Accel. Conf. (NAPAC 2022), Albuquerque, NM, USA (JACoW, 2022), pp. 249–252, Paper MOPA86. 10.18429/JACoW-NAPAC2022-MOPA86

[23] C. Compton *et al.*, "Fabrication efforts toward a superconducting RF photoinjector quarter-wave cavity for use in low-emittance injector applications," in Proc. 21st Int. Conf. RF Supercond. (SRF 2023), Grand Rapids, MI, USA (JACoW, 2023), Paper TUPTB063. 10.18429/JACoW-SRF2023-TUPTB063

[24] J. Smedley *et al.*, "Status of the CW SRF gun development at FRIB for LCLS-II-HE," in Proc. 22nd Int. Conf. RF Supercond. (SRF 2025), Tokyo, Japan (JACoW, 2026), pp. 603–607, Paper THB04. 10.18429/jacow-srf2025-thb04

[25] T. Konomi, C. Adolphsen, S. Gatzmaga, W. Hartung, M. P. Kelly, S. H. Kim, J. W. Lewellen, M. S. Patil, J. T. Popielarski, K. Saito, *et al.*, "Design of the cathode stalk for the LCLS-II-HE low emittance injector," in Proc. 5th North American Part. Accel. Conf. (NAPAC 2022), Albuquerque, NM, USA (JACoW, 2022), pp. 253–255, Paper MOPA87. 10.18429/JACoW-NAPAC2022-MOPA87

[26] T. Konomi, W. Hartung, S. H. Kim, S. Miller, D. Morris, K. Saito, A. Taylor, Z. Yin, T. Xu, M. P. Kelly, *et al.*, "Design and tests of a cathode stalk for the LCLS-II-HE low emittance injector SRF gun," in Proc. 21st Int. Conf. RF Supercond. (SRF 2023), Grand Rapids, MI, USA (JACoW, 2023), pp. 589–592, Paper TUPTB069. 10.18429/JACoW-SRF2023-TUPTB069

[27] Z. Y. Yin, W. Hartung, S. H. Kim, T. Konomi, and T. Xu, "Evaluation of photocathode-port multipacting in the SRF photoinjector cryomodule for the LCLS-II high-energy upgrade," in Proc. 21st Int. Conf. RF Supercond. (SRF 2023), Grand Rapids, MI, USA (JACoW, 2023), Paper WEPWB113. 10.18429/JACoW-SRF2023-WEPWB113

[28] X. Du, C. Adolphsen, Y. Choi, D. Greene, J. Lewellen, J. Wenstrom, and T. Xu, "Design of an emittance-compensation superconducting magnet package for LCLS-II-HE's SRF photoinjector," IEEE Trans. Appl. Supercond. **33**, 3500604 (2023). 10.1109/TASC.2023.3247699

[29] R. Xiang, A. Arnold, S. Gatzmaga, A. Hoffmann, P. Murcek, R. Steinbrück, J. Teichert, C. Adolphsen, J. Smedley, W. Hartung, *et al.*, "Design of a cathode insertion and transfer system for the LCLS-II-HE SRF gun," in Proc. 21st Int. Conf. RF Supercond. (SRF 2023), Grand Rapids, MI, USA (JACoW, 2023), pp. 267–270, Paper MOPMB067. 10.18429/JACoW-SRF2023-MOPMB067

[30] R. Xiang *et al.*, "Cathode insertion and transfer system for the LCLS-II-HE SRF gun," presented at the 16th Int. Part. Accel. Conf. (IPAC 2025), Taipei, Taiwan, 2025, Paper TUPM044. https://indico.jacow.org/event/81/contributions/7712/

[31] R. Xiang *et al.*, "Development of a particle-free cathode loadlock system for SRF-Gun," in Proc. 17th Int. Part. Accel. Conf. Deauville, France (IPAC 2026) (JACoW, in press), Paper THP2133. https://indico.jacow.org/event/95/contributions/14216/attachments/2144/8064/THP2133_edited_version.pdf

[32] R. Xiang, A. Arnold, H. Buettig, D. Janssen, M. Justus, U. Lehnert, P. Michel, P. Murcek, A. Schamlott, *et al.*, "Cs2Te normal conducting photocathodes in the superconducting rf gun," Phys. Rev. ST Accel. Beams **13**, 043501 (2010). 10.1103/PhysRevSTAB.13.043501

[33] J. Kühn, J. Borninkhof, M. Bürger, A. Frahm, G. Klemz, S. Lederer, A. Neumann, H. Pflocksch, M. Schuster, *et al.*, "UHV photocathode plug transfer chain for the bERLinPro SRF-photoinjector," in Proc. 8th Int. Part. Accel. Conf. (IPAC 2017), Copenhagen, Denmark (JACoW, 2017), pp. 1381–1383, Paper TUPAB029. 10.18429/JACoW-IPAC2017-TUPAB029

[34] J. Bisognano *et al.*, "Wisconsin SRF Electron Gun Commissioning," in Proc. North American Part. Accel. Conf. (NAPAC 2013), Pasadena, CA, USA (JACoW, 2013), Paper TUPMA19, pp. 622–624. https://jacow.org/PAC2013/papers/tupma19.pdf

[35] J. R. Harris, K. L. Ferguson, J. W. Lewellen, S. P. Niles, B. Rusnak, R. L. Swent, W. B. Colson, T. I. Smith, C. H. Boulware, T. L. Grimm, *et al.*, "Design and operation of a superconducting

quarter-wave electron gun," Phys. Rev. ST Accel. Beams **14**, 053501 (2011). 10.1103/PhysRevSTAB.14.053501

[36] J. Dube, C. Wang, E. J. Brookes, A. Neumann, G. Klemz, E. Ergenlik, P. Echevarria, A. Ushakov, T. Kamps, and J. Kühn, "Quantitative framework for photoemissive materials: From bialkali-antimonide synthesis to performance in superconducting RF electron injectors," Phys. Rev. Mater. **10**, 074601 (2026). 10.1103/d2s8-7pt8

[37] F. Sannibale, D. Filippetto, H. Qian, C. Mitchell, F. Zhou, T. Vecchione, R. K. Li, S. Gierman, and J. Schmerge, "High-brightness beam tests of the very-high-frequency gun at the Advanced Photoinjector Experiment test facility at Lawrence Berkeley National Laboratory," Rev. Sci. Instrum. **90**, 033304 (2019). 10.1063/1.5088521

[38] L. Wang, A. Brachmann, F.-J. Decker, Z. Li, T. O. Raubenheimer, J. F. Schmerge, and F. Zhou, "Cathode ion bombardment in LCLS and LCLS-II RF gun," in Proc. 37th Int. Free Electron Laser Conf. (FEL 2015), Daejeon, Korea (JACoW, 2015), pp. 534–539, Paper TUP072. 10.18429/JACoW-FEL2015-TUP072

[39] F. Zhou, C. Adolphsen, A. Benwell, G. Brown, D. H. Dowell, M. Dunning, K. Grouev, G. Hays, J. Hodgkinson, T. M. Huang, *et al.*, "Commissioning of the SLAC Linac Coherent Light Source II electron source," Phys. Rev. Accel. Beams **24**, 073401 (2021). 10.1103/PhysRevAccelBeams.24.073401

[40] F. Zhou, G. Adhikari, A. Brachmann, D. Cesar, W. Colocho, Y. Ding, D. Dowell, G. Just, D. Kennedy, D. Magana, et al., "Linac Coherent Light Source II injector commissioning and early operational experience," Phys. Rev. Accel. Beams **29**, 104501 (2026). 10.1103/1wk3-p29y

[41] L. Zheng, H. Chen, B. Gao, Z. Dong, Z. Li, Y. Jia, Q. Tian, Q. Xia, Y. Zhu, J. You, *et al.*, "Design, fabrication, and beam commissioning of a 216.667 MHz continuous-wave photocathode very-high-frequency electron gun," Phys. Rev. Accel. Beams **26**, 103402 (2023). 10.1103/PhysRevAccelBeams.26.103402

[42] X.-H. Wang, G. Shu, H.-J. Qian, X.-D. Li, Z.-P. Liu, Z.-G. Jiang, H. Meng, C.-C. Xing, Q. Zhou, and H.-X. Deng, "Experimental demonstration of dark current mitigation by an over-inserted plug in a normal conducting very-high-frequency gun," Phys. Rev. Accel. Beams **28**, 043401 (2025). 10.1103/PhysRevAccelBeams.28.043401